\documentclass[twocolumn,showpacs,preprintnumbers,amsmath,amssymb]{revtex4}

\usepackage{graphicx}
\usepackage{dcolumn}
\usepackage{bm}

\begin{document}

\title{Backward Raman scattering in relativistic electron beam and 
intense THz light}

\author{S. Son}
\affiliation{169 Snowden Lane, Princeton, NJ 08540}

\begin{abstract}
A new type of the THz laser is proposed. A coherent tera-hertz light is emitted through  the backward Raman scattering  between a visible light laser and a relativistic  electron beam. The threshold conditions for the laser intensity and the electron beam density are identified. This scheme may lead to one of the most intense tera-hertz coherent light sources.
\end{abstract}

\pacs{42.55.Vc, 42.65.Ky, 52.38.-r, 52.35.Hr}       

\maketitle
 An intense and compact  THz light source is critical component  in the biomedical image, the tomography, the molecular spectroscopy, the tele-communication and many others~\cite{siegel,siegel2,  siegel3, booske,radar, diagnostic, security}.
Many THz light sources have been invented~\cite{Tilborg,Zheng,Reimann, gyrotron, gyrotron2, gyrotron3,magnetron, qlaser, qlaser3, freelaser, freelaser2, colson, songamma,Gallardo,sonttera}, but 
even the most advanced ones are neither practical nor intense enough; 
 The free electron laser~\cite{ freelaser, freelaser2, colson} needs expensive magnets and accelerators, the quantum cascade laser~\cite{ qlaser, qlaser3} cannot produce intense THz light neither be operated in the room temperature and the gyrotron suffers the scale problem unless the magnetic field is ultra intense~\cite{ gyrotron, gyrotron2, gyrotron3,magnetron}. The current inability to produce the power (intensity) high enough for  various applications aforementioned is referred as ``THz gap''. 

In this paper, the author proposes a new process of the THz light generation 
by amplifying the interaction between an intense visible-light laser and a relativistic electron beam via the backward Raman scattering (BRS).
To my  knowledge, it is the first scheme, in which  a  visible light wave is shifted down into THz light via the BRS, potentially overcoming various difficulties of the current technoloiges.
In the visible-light laser compression~\cite{Fisch} and the inertial confinement fusion research~\cite{tabak}, 
the BRS  has been demonstrated to be much stronger  than the scattering by an individual electron; 
As an illustration, considering the electron plasma with the density $10^{16} $ particles per cubic centimeter and the visible-light laser with the intensity of $10^{15}$ watt per square centimeter, the BRS is $10^{11}$ times larger than the conventional Thomson scattering.
 If utilized produently, a light source based this strong scattering can exceed the free electron laser by order of magnitude, which is the main motiviation of this work. 
It is worthwhile to point out that so far, the most powerful current THz light source is based on the Thomson scattering. 

\begin{figure}
\scalebox{0.3}{
\includegraphics{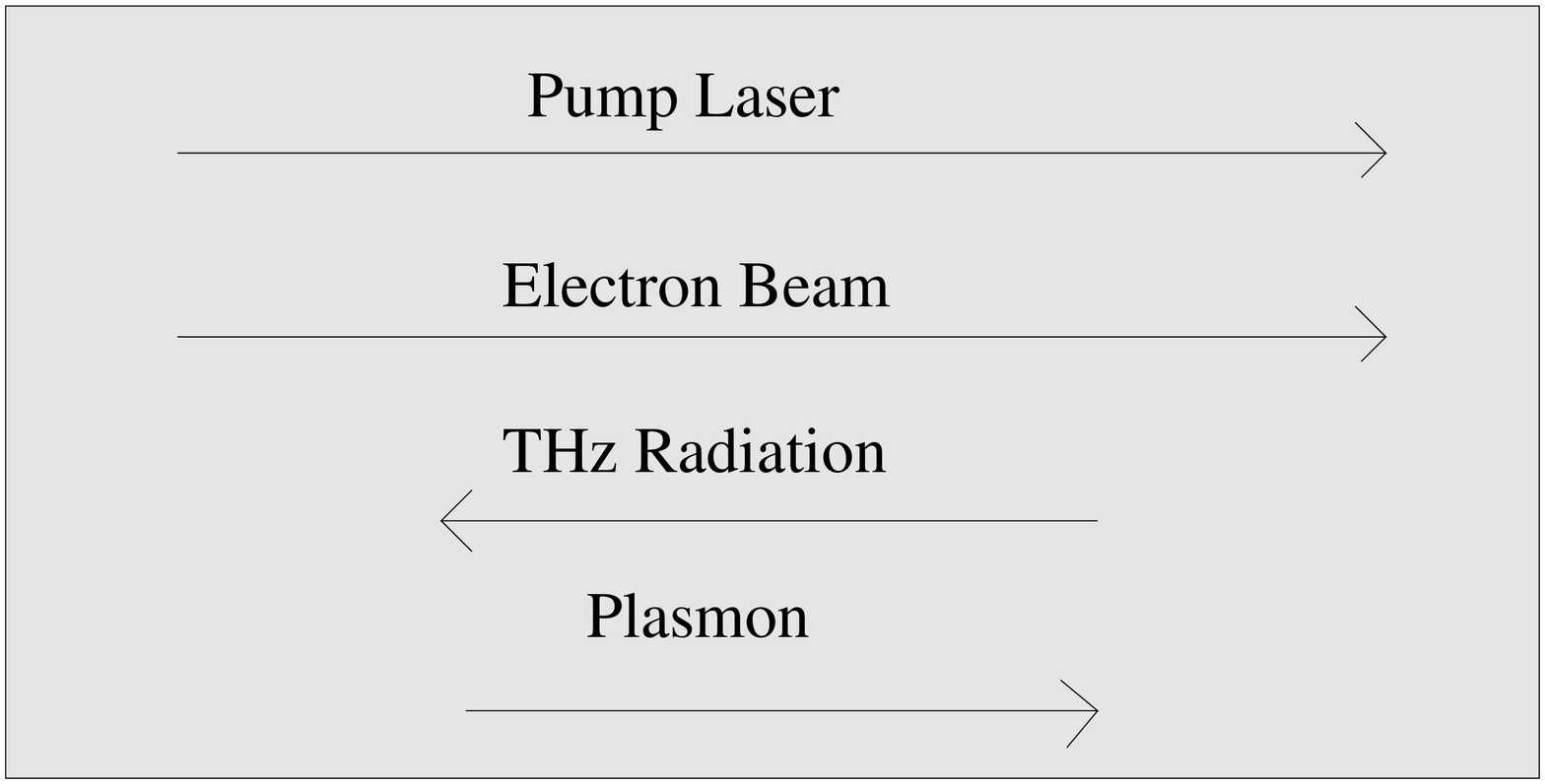}}
\caption{\label{fig:1}
The schematic diagram  about the propagation direction of the BRS pump, THz pulse, the electron beam and the plasmon. 
}
\end{figure}

\begin{figure}
\scalebox{1.0}{
\includegraphics[width=0.7\columnwidth, angle=270]{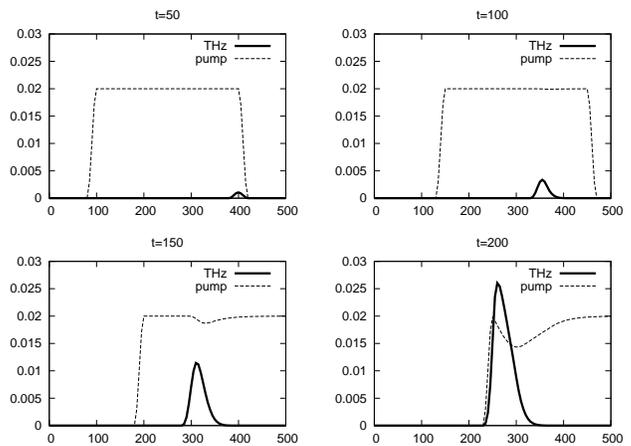}}
\caption{\label{fig:2}
The 1-D simulation of the BRS in the co-traveling frame, where $n_0 = 10^{16} \ \mathrm{cm^{-3}} $ and $\gamma_0=3.0$. 
The $x$-axis is normalized by $x_c^{-1} = \omega_{\mathrm{pe}}/c\sqrt{\gamma_0}$  ($x_c \cong 0.0075\  \mathrm{cm} $) and the y-axis is the quiver velocity divided by the velocity of the light ($\mathrm{A} = A_p$ or $\mathrm{A}= A_s$), as defined in Eq.~(\ref{eq:2}).
Initially ($t=0$), the THz pulse is located at $x=450 \ x_c $ with the peak of $\mathrm{A} = 0.001$ corresponding to 
$I = 1.5 \times 10^{7} \ \mathrm{W} / \mathrm{cm}^2 $.  The inital pump laser  is located between $x=0 $ and $x=350 \  x_c $, with  the peak of $\mathrm{A} = 0.02$ corresponding to $I= 6 \times 10^{12} \ \mathrm{W} / \mathrm{cm}^2 $.  The pump pulse is the CO2 laser with $\lambda_{s0} = 10 \ \mu \mathrm{m}$.
The pump and seed pulses has the duration of 20 pico-seconds in the laboratory frame.  The pump (xray) propagetes from the right (left) to the left (right).
 The top-left is the initial THz  pulse and pump at $t= 50 \ \tau_c $ where $\tau_c = 0.25$ pico-second.  
   The top-right is the THz pulse and the pump at $t=100 \ \tau_c$.   The bottom-left is the THz pulse and the pump at $t=150 \ \tau_c$. The bottom-right is the THz pulse and the pump at $t=300 \ \tau_c$. 
The peak intensity of the THz, which is proportional to $\mathrm{A}^2$  has been amplified by a factor of 1000. 
}
\end{figure}

   The author considers the situation when an intense visible light laser \textit{co-traveling} with a relativistic electron beam excites a Langmuir wave and  the BRS between the Langmuir wave and the laser emits a THz light \textit{in the opposite direction} to the electron beam (Fig.~\ref{fig:1}). In this case, the laser will be shifed down to the THz light via the relativistic Doppler's effect.  The analysis in this paper suggests that the scheme cold produce the most powerful and efficient THz light source and  even possibly a  cheap one.  
The linear analysis of the 1-D BRS is provided analytically and the range of the physical parameters (Table~\ref{tb}), at which the current scheme becomes most practical, are estimated. 
  One-dimensional (1-D) simulation of the pulse 
amplification (Fig.~\ref{fig:2}) is performed to validate its plausibility. 


To begin with, consider a relativistic electron beam and a co-traveling laser (pump laser). Let us denote the beam  density in the laboratory frame by $n_0$ and the beam relativistic factor by $\gamma_0= (1-v_0^2/c^2)^{-1/2}$, where $v_0$ is the velocity of the electron beam. In the co-traveling frame, the electron density becomes $n_1 = n_0 / \gamma_0$ due to the length dilation. The BRS between the laser (the pump pulse) and the electron beam could emit the THz light (the seed pulse). The 1-D BRS three-wave interaction in the \textit{co-traveling} frame is~\cite{McKinstrie}:
\begin{eqnarray}
\left( \frac{\partial }{\partial t} + v_p \frac{\partial}{\partial x} + \nu_1\right)A_p  = -ic_p A_s A_3  \nonumber \mathrm{,}\\
\left( \frac{\partial }{\partial t} + v_s \frac{\partial}{\partial x} + \nu_2\right)A_s  = -ic_s A_p A^*_3   \label{eq:2} \mathrm{,} \\
\left( \frac{\partial }{\partial t} + v_3 \frac{\partial}{\partial x} + \nu_3\right)A_3  = -ic_3 A_p A^*_s  
\nonumber \mathrm{,}
\end{eqnarray}
where $A_i= eE_{i1}/m_e\omega_{i1}c$ is the ratio of  the electron quiver velocity of the pump pulse ($i=p$) and the seed pulse ($i=s$) relative to the speed of light $c$, $E_{i1}$ is the electric field of the seed (pump) pulse, $A_3 = \delta n_1/n_1$ is the Langmuir wave amplitude, $\nu_1 $ ($\nu_2$) is the inverse bremsstrahlung rate of the pump (seed), $\nu_3$ is the plasmon decay rate, $ c_i = \omega_3^2/ 2 \omega_{i1}$ for $i=p, s$, $c_3 = (ck_3)^2/2\omega_3$, $ \omega_{p1} $ ($ \omega_{s1} $) is the frequency of the pump (seed) laser and $\omega_3 \cong \omega_{pe} / \sqrt{\gamma_0} $ is the plasmon  frequency. In the co-traveling frame, the wave vector (frequency) of a photon satisfies the usual dispersion relation, $\omega_1^2 =   \omega_{pe}^2/\gamma_0+ c^2 k_1^2$, where $\omega_1$ and $k_1$ are the photon wave frequency and the corresponding vector, and $\omega_{pe}^2 = 4 \pi n_0 e^2 /m_e$ is the plasmon frequency. Denote the wave vector and the frequency of the pump laser (seed pulse or THz light) in the co-traveling frame as $k_{p1}$ and $\omega_{p1} $ ($k_{s1} $ and $\omega_{s1} $) and their laboratory frame counterparts as $k_{p0}$ and $\omega_{p0} $ ($k_{s0} $ and  $\omega_{s0} $). The Lorentz transformation leads to the following relationship:
\begin{eqnarray} 
\omega_{p0} &=& \gamma_0 \left[ \sqrt{\omega_{pe}^2/\gamma_0 + c^2 k_{p1}^2 } + vk_{p1} \right] \mathrm{,}  \label{eq:lorentz1} \\  \nonumber \\
k_{p0} &=&  \gamma_0 \left[ k_{p1} + \frac{\omega_{p1} }{c}  \frac{v_0}{c} \right] \mathrm{,} \label{eq:lorentz2} \\ \nonumber \\ 
 \omega_{s0} &=& \gamma_0 \left[ \sqrt{\omega_{pe}^2/\gamma_0 + c^2 k_{s1}^2 } - 
vk_{s1} \right] \mathrm{,}  \label{eq:lorentz3} \\  \nonumber \\ 
k_{s0} &=&  \gamma_0 \left[ k_{s1} - \frac{\omega_{s1} }{c}  \frac{v_0}{c} \right]\mathrm{.} \label{eq:lorentz4} \\ \nonumber 
\end{eqnarray}
The energy and momentum conservation of Eq.~(\ref{eq:2}) are given as 
\begin{eqnarray} 
 \omega_{p1} &=& \omega_{s1} + \omega_{3} 
\mathrm{,} \nonumber \\ 
 k_{p1} &=&  k_{s1} +k_3 \mathrm{,} \label{eq:cons} 
\end{eqnarray}
where $k_3$ is the wave vector of the plasmon. With a given pump frequency $\omega_{p0} $, $k_{p1} $ ($\omega_{p1} $) is determined from Eq.~(\ref{eq:lorentz1}), $k_{s1} $ ($\omega_{s1}$) is determined from Eq.~(\ref{eq:cons}), and, finally,  $k_{s0} $ ($\omega_{s0}$) is determined from  Eqs.~(\ref{eq:lorentz3}) and (\ref{eq:lorentz4}). In the limiting case $ck_{s1} \gg \omega_3 $, $\omega_{s0} \cong (1/2 \gamma_0) (\omega_{p1} - \omega_3)$ or   
\begin{equation} 
\omega_{s0} \cong \frac{1}{4\gamma_0^2} \left[ \omega_{p0} - 2 \omega_{pe} \sqrt{\gamma_0}\right] \mathrm{,}\label{eq:down}
\end{equation} 
where $\omega_{p1} \cong  \omega_{p0} / 2 \gamma_0 $ and $\omega_3 \cong \omega_{pe} /\sqrt{\gamma_0} $. Eq.~(\ref{eq:down}) describes the frequency down-shift of the visible-light pump laser into the THz light, via the relativistic Doppler effect. For instance, if $\gamma_0= 3$, the down-shifted frequency would be $0.8$ THz for the CO2 laser whose frequency is  30 THz.

The BRS growth rate can be obtained  from Eq.~(\ref{eq:2}). When $|A_p| \gg |A_s|$  and $|A_p| \gg |A_3|$, the linearization of Eq.~(\ref{eq:2}), in the expansion of $A_{s, 3}(t) = A_{s,3}(\omega) \exp(i\omega t) $, leads to 
\begin{eqnarray} 
\omega^2 &+& (\nu_3 + \nu_2) \omega + (\nu_3 \nu_2) - c_s c_3 |A_{p}|^2=0  \mathrm{,} 
\label{eq:inst} 
\end{eqnarray}  
If $\nu_2 =0$ and $\nu_3 \ll ( c k_{p1} \omega_3)^{1/2} |A_{p}|$, the growth rate, the imaginary part of the solution of Eq.~(\ref{eq:inst}), is $\Gamma_1 \cong  \sqrt{ c_s c_3 } |A_p|  $. In the limiting case $ck_{p} \gg \omega_3 $, the Lorentz transformation prescribes $E_{p1} / E_{p0} \cong 1/2 \gamma_0$ and $A_{p} = eE_{p1}/m_e\omega_{p1}c \cong  eE_{p0}/m_e\omega_{p0}c$, resulting in
\begin{equation} 
\Gamma_1 \cong \sqrt{\omega_{pe} \omega_{p0} /2\gamma_0^{3/2}} |A_{p}| \mathrm{.} \label{eq:inst2}
\end{equation}
Denoting the electron beam length as $L_b$ and the laser length as $L_l$, the beam length in the co-traveling frame increases to $\gamma_0 L_b $ and the laser length increases to $2 \gamma_0 L_l $ so that the interaction time between the beam and the laser is $ \tau \cong \min(\gamma_0 L_b, 2 \gamma_0 L_l)/c $. The gain-per-length $g$ is therefore estimated as
\begin{eqnarray} 
g &=& \Gamma_1 \tau / L_b = \gamma_0 \Gamma_1 /c \ \ \  \mathrm{if}   \ L_b < 2 L_l  \mathrm{,} \nonumber \\ 
g &=& \Gamma_1 \tau / L_l = 2\gamma_0 \Gamma_1 /c \ \  \mathrm{otherwise.} \label{eq:gain}   
\end{eqnarray}

\begin{table}[t]
\centering
\begin{tabular}{|c||cccccccc||}
	\hline
  Type &  \textbf{Freq}    &   $I_{15}$   &  $\gamma_0$ &  $A_1$ & $\Gamma_{13}$ & $g_1$ & $g_2$ &  $n_c$   \\
	\hline \hline 
N &    \textbf{3.0} &   1.00 &    5.0  &      0.020 &       0.0245 &    40.874  &     81.748  &     2.56 \\ 
N   &  \textbf{3.0}  &  $10^{3}$ &  5.0  &     0.6324   &    0.775  &    1292  &    2585  &   2.56 \\
N   &  \textbf{3.0}  &    0.01  &   5.0  &     0.0006  &    0.0008 &  1.29  &    2.58    &   2.56 \\
N  &   \textbf{8.3} &    1.0  &    3.0  &      0.02   &      0.0359  &   35.97 &       71.95 &       7.11 \\
N   &  \textbf{8.3}   & $10^3$ &     3.0    &    0.63 &        1.14 &     1137  &    2275 &      7.1 \\
N  &    \textbf{8.3} &    0.01  &   3.0    &    0.002 &      0.0036 &   3.597 &       7.19  &      7.1 \\
C   &  \textbf{0.83} &   1.0&     3.0&      0.2  &       0.113 &     113.75&       227.5 &      0.007 \\
C   &  \textbf{0.83}&    30&    3.0&      1.0 &         0.62&     623&     1246&    0.007 \\
C    & \textbf{0.83}  &  $10^{-3}$ &   3.0  &    0.006 &      0.0036 &   3.59&       7.2 &       0.007 \\ 
C   &  \textbf{1.88} &  1.0 &     2.0&       0.2 &        0.15&    103 &     206 &      0.016 \\
C   &  \textbf{1.88} &   0.01 &    2.0&       0.02&        0.015&    10.28 &      20.56 &     0.016 \\
\hline
\end{tabular}
\caption{ The laser and electron beam parameter and the characteristic of the THz radiation. \label{tb} In this example, I assume that $n_0 = 10^{15} \ / \mathrm{cc}$. In the table, N (C) stands for the NA:YAG laser (CO2 laser) with the wave length of $1 \ \mu \mathrm{m} $ ($10 \ \mu \mathrm{m} $), $I_{15}$ is the laser intensity normalized by $10^{15} \ \mathrm{W} / \mathrm{cm}^2$, $\gamma_0$ is the relativistic factor, $A_1$ is the quiver velocity divided by the velocity of light as defined in Eq.~(\ref{eq:2}), Freq is the seed pulse frequency $F = \omega_{s0} / 2 \pi $  as given in Eq.~(\ref{eq:down}) in the unit of $10^{12} \sec$, $\Gamma_{13} $ is the growth rate normalized by $10^{13} / \sec$ as given in Eq.~(\ref{eq:inst2}),  $g_1$ ($g_2$) is the gain-per-length from Eq.~(\ref{eq:gain}) in the unit of $\mathrm{cm}^{-1} $ and $n_c$ is the lower bound of the density given in Eq.~(\ref{eq:cond}) normalized by $10^{12}\  /\mathrm{cc} $. 
}
\end{table}

An estimation of the energy conversion efficiency of the pump to the seed pulse is as follows. Practical applications of the BRS compression of the visible-light lasers have demonstrated that a significant portion of the pump energy can be converted to the seed pulse~\cite{malkin1, Fisch, BBRS, BBRS2, BBRS3, sonforward}. Denote  the total energy of the pump laser by $\mathrm{E}_{p0}$, which becomes $\mathrm{E}_{p1} = \mathrm{E}_{p0}/2 \gamma_0$ in the co-traveling frame, and denote the conversion efficiency in the co-traveling frame by $\epsilon_1 $. Then,  the energy transferred from the pump to the seed is $\mathrm{E}_{s1} = \epsilon_1 \mathrm{E}_{p0}/2 \gamma_0$, which is $\mathrm{E}_{s0} = \epsilon_1 \mathrm{E}_{p0}/4 \gamma_0^2$ in the laboratory frame. Therefore, the conversion efficiency in the laboratory frame is 
\begin{equation} 
\epsilon_0 \cong \epsilon_1 / 4 \gamma_0^2 \mathrm{.}  \label{eq:eff}
\end{equation}
The CO2 laser has the wavelength of $10  \ \mu \mathrm{m} $ and the  Nd:YAG laser has the wavelength of $1 \ \mu \mathrm{m} $. From Eq.~(\ref{eq:down}), for a fixed $\omega_{s0} $, the required relativistic factor for the CO2 laser should be lower than the Nd:YAG laser by a factor of $\sqrt{10}$ and thus the conversion efficiency $\epsilon_0 $ for the CO2 laser will be larger than the Nd:YAG laser by a factor 10 for the same $\epsilon_1$. For $\gamma\cong 3$, the conversion efficiency for the CO2 laser can be as high as a few percents, assuming that $\epsilon_1$ is a few tens of percents.

In order for the current scheme to work, there exist a few necessary conditions for the laser and the electron beam. One necessary condition is $  \Gamma_1 \tau > 1 $ for a sufficient amplification or  
\begin{equation} 
 |A_{p}|  > \frac{2^{1/2}}{\gamma_0^{1/4}} \frac{c}{\mathrm{min}(L_l, L_b) } \left( \frac{1}{ \omega_{pe}\omega_{p0}} \right)^{1/2}  \label{eq:laser} \mathrm{,}
\end{equation} 
which is  readily satisfied by currently available intense visible-light laser~\cite{cpa, cpa2, cpa3,cpa4}. Another  necessary condition  for the electron beam density is $n_0 / \gamma_0 >  k_3^3 / (2 \pi)^3 $, as only the plasmons  with $n_0 /\gamma_0 \gg k_3^2 $ are collective waves. In the limiting case $ c k_{p1} \gg \omega_3$, the condition is  
\begin{equation} 
 n_0 >     \gamma_0^{-2} \left(\frac{k_{p0}^3}{  8\pi^3 }  \right)
 \cong 64\times  \gamma_0^{4} \left(\frac{k_{s0}^3}{  8\pi^3 }  \right)
   \label{eq:cond}  \mathrm{,}
\end{equation} 
where $k_{p1} \cong k_{s1} \cong k_{p0} /  2 \gamma_0$ and $k_{s0} \cong k_{p0} / 4 \gamma_0^2$. In addition, the wave vector of the Langmuir wave should be larger than the Debye length

\begin{equation}
k_3 < \lambda_{de}^{-1}   \label{eq:cond2}
\end{equation} 
where $\lambda_{de}^{-2} =  4 \pi n_0 e^2 / \gamma_0 m_e $. The condition given by Eqs.~(\ref{eq:cond}) and (\ref{eq:cond2}) estimates the minimum electron beam density for the BRS compression for a given electron temperature. Note that much higher density can be achieved through the current electron accelerator or dense electron beams~\cite{monoelectron, ebeam, tabak,sonprl,sonpla} while the electron temperature is low enough to satisfy Eq.~(\ref{eq:cond2}).

The estimations  for various electron beams and lasers are provided in the Table~\ref{tb}. The Table  suggests that the proposed scheme is plausible for a wide range of frequencies, the growth rate (the gain-per-length) can be as high as $ 10^{13} / \sec $ ($10^3 / \mathrm{cm} $) and the requirement of the laser intensity, $I > 10^{12} \ \mathrm{W} / \mathrm{cm}$, is moderate. If $I_{15} \geq 10^3 $ for the case of the Nd:YAG laser as in Table~\ref{tb}, the electron quiver velocity becomes marginally relativistic and the full relativistic treatment is necessary~\cite{brs}, which is ignored in this paper. 

In Fig.~(\ref{fig:2}),
 a  1-D simulation of Eq.~(\ref{eq:2}) is performed, where the pump laser is the CO2 laser with $I = 6 \times 10^{12} \ \mathrm{W} / \mathrm{cm}^2 $ ($A_p = 0.02$). The pump laser (the seed pulse)  moves from the left (right) to the right (left) and the THz pulse extracts the energy from the pump laser via the BRS, resulting the energy gain of  the THz pulse  by a factor of 1000. 
In this example,  the pump laser has the intensity of  $I= 1.5 \times 10^{12} \ \mathrm{W} / \mathrm{cm}^2 $ ($I=  6 \times 10^{12} \ \mathrm{W} / \mathrm{cm}^2 $) in the co-moving frame (the laboratory frame);   
the  intensity of the pump laser (THz pulse) is higher (lower)  by 36 times in the laboratory frame  than in the co-moving frame  due to the Doppler's effect. 
In the simulation shown in Fig.~(\ref{fig:2}), the final peak intensity of the THz pulse  in the co-moving frame  is comparable to the intensity of the pump laser; 
 the attained peak intensity of the THz pulse  is $I= 1.1 \times 10^{10} \ \mathrm{W} / \mathrm{cm}^2 $ in the laboratory frame, which is 0.1 percent of the laser pump intensity.  Various simulation suggests that 
the  attained peak intensity of the THz pulse  in the laboratory frame can be as high as the the laser pump intensity.  

Fig.~(\ref{fig:2}) illustrates that the final peak intensity of the THz light can be very high; The THz light with such high intensity can be very useful for the spectroscope and the dynamic imaging application, noting the fact that the intensity from the current availble light sources is at least million times smaller.

To summarize, a new scheme of the THz light source is proposed based the backward Raman scattering. 
 Detailed estimation of the optimal parameter range for the laser beam and the electron beam are provide in Table~\ref{tb} and one example on the 1-D simulation of Eq.~(\ref{eq:2}) is provided in Fig.~(\ref{fig:2}). The most intense and powerful coherent THz light source can be manufactured based on the current scheme; The gain-per-length can be as high as 1000 per centimeter while the current strongest THz laser has the gain-per-length much less than 1 per centimeter, 
the conversion efficiency of converting the input laser energy into the THz light is as high as a few percents while the current most efficient light source has the conversion efficiency less than 0.01 percent,  the THz light energy per one laser shot can be as high as a few J while the most powerful THz laser can produce the enery-per-shot much lower than 0.001 J and, finally, 
the peak THz intensity can be comparable the pump laser intensity.  In addition, because it does not need expensive magnets or high quality electron beam, the disclosed light source has advantages in the compactness, the mobility and the operating (construction) cost. 
While some researches have attempted to  shift the visible light laser up to the XUV regime via the BRS~\cite{brs, brs2, brs3}, the new idea for the scheme presented here  is that the visible-light laser is  \textit{shifted down} to the THz regime via the BRS.


There are technical issues to consider in the realization of the proposed scheme. The performance of the proposed scheme relies on the quality of the electron beam such as the uniformity and the time duration in which the beam maintains its quality. Also, a rather intense seed THz pulse would be needed in order to extract the significant fraction of the pump pulse energy. However, even if the difficulties mentioned compromises the efficiency of the scheme proposed to some degree, its prospect as a powerful THz light source remains high. 

\bibliography{tera2}

\end{document}